\documentclass{webofc}
\usepackage[varg]{txfonts}   
\newcommand{\vect}[1]{\boldsymbol{#1}_{\perp}}
\newcommand{\abar}{\bar{\alpha}_s}
\newcommand{\qhat}{\hat{q}}
\newcommand{\kt}{\vect{k}}

\newcommand{\rmd}{\textrm{d}}

\newcommand{\qt}{\vect{q}}

\newcommand{\rt}{\vect{r}}

\begin{document}
\title{Anomalous dimension of transverse momentum broadening in planar $\mathcal{N}=4$ SYM}
%
%

\author{\firstname{Paul} \lastname{Caucal}\inst{1}\fnsep\thanks{\email{caucal@subatech.in2p3.fr}} 
}

\institute{SUBATECH UMR 6457 (IMT Atlantique, Universit\'e de Nantes, IN2P3/CNRS), 4 rue Alfred Kastler, 44307 Nantes, France
          }

\abstract{%
  The typical transverse momentum $Q_s(t)$ (or "saturation" momentum) acquired by a hard particle propagating through a $\mathcal{N}=4$ SYM plasma increases over time like $t^\gamma$, with an anomalous exponent $\gamma>1/2$ characteristic of super-diffusion. This anomalous exponent is a function of the 't Hooft coupling $\lambda=g^2N_c$. Recently, a method has been proposed to systematically compute the perturbative series of $\gamma(\lambda)$ at weak coupling. This method relies on the traveling wave interpretation of the time evolution of $Q_s(t)$ and on the dominance of soft-collinear radiative corrections at large times. In this paper, we compute $\gamma(\lambda)$ up to $\mathcal{O}(\lambda^{2})$ using the double logarithmic behaviour of the BFKL equation in planar $\mathcal{N}=4$ SYM at three loops. This calculation allows us to discuss the transition towards the strong coupling regime where AdS/CFT calculations predict $\gamma\to 1$.
}
\maketitle
\section{Introduction}
\label{intro}

High energy partons propagating through a hot and dense QCD medium suffer multiple soft scatterings with the medium quasi-particles that increase their transverse momentum with respect to their direction of motion. This phenomenon, known as transverse momentum broadening (TMB), is quantified by the jet quenching parameter $\qhat$  which is typically the average transverse momentum squared acquired per unit of time $t$ in the plasma, $\langle k_\perp^2\rangle \sim \qhat t$. It is a fundamental ingredient of the jet quenching effect measured both at RHIC and at the LHC \cite{Blaizot:2015lma,Qin:2015srf,Cunqueiro:2021wls}. For instance, it leads to the transport at large angles of the deposited energy by high energy jets \cite{Caucal:2019uvr,Mehtar-Tani:2021fud} through a turbulent cascade \cite{Blaizot:2013hx,Blaizot:2013vha} that efficiently degrades the initial energy down to thermal scales \cite{Iancu:2015uja,Schlichting:2020lef,Mehtar-Tani:2022zwf}. 

From a more theoretical perspective, several recent studies have taken a fresh look at TMB beyond leading order in perturbative QCD (pQCD) \cite{Caucal:2021lgf,Arnold:2021mow,Arnold:2021pin,Caucal:2022fhc,Ghiglieri:2022gyv,Caucal:2022mpp}. At NLO, the recoil of the highly energetic parton from the gluon emission contributes to the increase of its transverse momentum. The quantum radiative corrections to $\qhat$ have been computed for the first time in the seminal paper \cite{Liou:2013qya}. The authors have shown that the leading NLO corrections are double logarithmically enhanced like $\propto\alpha_s \ln^2(L/\tau_0)$ where $L$ is the medium size and $\tau_0\ll L$ is a thermal scale. The presence of a double logarithm suggests that the underlying physical mechanism is the emission of soft and collinear gluons. As we shall see, the numerical prefactor comes from the allowed phase space for these emissions which is constrained by multiple soft scattering effects. One already notices that this problem is intrinsically \textit{non-linear} since the constraint put on the radiations depends on $\qhat$ itself.

This logarithmic enhancement turns out to be independent of the physical process where $\qhat$ is involved and can therefore be absorbed into a "renormalized" quenching parameter \cite{Blaizot:2014bha,Iancu:2014kga,Arnold:2021mow,Arnold:2021pin}. We emphasize that these corrections are distinct from the $\mathcal{O}(g)$ corrections $\qhat$ computed in \cite{Arnold:2008vd,Caron-Huot:2008zna} in pQCD and in \cite{Ghiglieri:2018ltw} in $\mathcal{N}=4$ SYM. As noted in \cite{Ghiglieri:2022gyv}, these corrections smoothly match the double logarithmic ones in the soft phase space where the gluon energy is between $gT$ and $T$, the thermal scales.

The large NLO correction to $\qhat$ questions the validity of the perturbative approach to get accurate predictions for the transport coefficient $\qhat$, unless a resummation is performed to all orders in perturbation theory. This resummation has been done in the double logarithmic and linear approximation in \cite{Liou:2013qya,Blaizot:2014bha,Iancu:2014kga}. The system size dependence is dramatically modified since one finds $\langle k_\perp^2\rangle \propto L^{1+2\sqrt{\abar}}$ with $\abar=\alpha_sN_c/\pi$. This anomalous scaling signals the onset of a super-diffusive regime induced by the non-locality of quantum corrections \cite{Caucal:2021lgf}. As outlined in \cite{Iancu:2014kga,Blaizot:2014bha}, it is interesting to notice that this scaling behaviour lies between the $\alpha_s\to 0$ limit and the strong coupling limit $\langle k_\perp^2\rangle \propto L^2$ obtained from the AdS/CFT correspondance \cite{Hatta:2007cs,Hatta:2008tx,Dominguez:2008vd}. 

In this paper, we compute the next four terms in the development of the scaling exponent in planar $\mathcal{N}=4$ SYM and discuss in more details the transition towards the strong coupling regime. 
We consider the conformal planar $\mathcal{N}=4$ SYM theory to exploit strong coupling results from AdS/CFT correspondence and the knowledge of the BFKL equation \cite{Kuraev:1977fs,Balitsky:1978ic} at three loops. The case of QCD with massless quarks and in the large $N_c$ limit has been covered in \cite{Caucal:2022mpp}. Conformal symmetry breaking through the running of $\alpha_s$ changes the behaviour of the scaling exponent into $1+4(\beta_0\ln(L/\tau_0))^{-1/2}$ where $\beta_0$ is the one-loop coefficient of the QCD $\beta$-function. We point out that we will often take the liberty to switch from fixed coupling QCD to $\mathcal{N}=4$ SYM whenever the correspondance is obvious from the relation $\abar \to \lambda/(4\pi^2)$.

We rely on the mapping \cite{Caucal:2021lgf,Caucal:2022fhc} between the evolution equation that resums the logarithmically enhanced radiative corrections and the equation that governs reaction-diffusion processes  with a traveling wave interpretation \cite{dee1983propagating,van1987dynamical,Munier:2003vc,2000,2003}. The most important consequence of this mapping is the universal property of the scaling limit which does not depend neither on the non-perturbative input for the initial condition nor on the "details" of the multiple soft scattering boundary \cite{Caucal:2022fhc}. In the end, the double logarithmic behaviour of the evolution kernel is sufficient to get the weak coupling development of the anomalous dimension.
\section{Non-linear evolution of transverse momentum broadening}
\label{sec:1}

We start by fixing our notations and conventions. The TMB distribution is defined as the probability distribution for a hard particle propagating through the dense medium to acquire a transverse momentum $\kt$ over a distance $L$. This distribution is encoded in the forward scattering amplitude $\mathcal{S}(\rt)$ of an effective dipole in the adjoint representation via a Fourier transform
\begin{equation}
\mathcal{P}(\kt)=\int\rmd^2\rt e^{-i\kt\cdot\rt}\mathcal{S}(\rt)\,.
\end{equation}
 For local and instantaneous interactions, $\mathcal{S}(\rt)$ exponentiates
\begin{equation}
\mathcal{S}(\rt)=e^{-\frac{1}{4}\sigma_{\rm dip}(\rt)L}\,,\quad\sigma_{\rm dip}(\rt)=\int\frac{\rmd^2\qt}{(2\pi)^2}\left(1-e^{i\qt\cdot\rt}\right)\mathcal{C}(\qt)\,,\label{eq:S-matrix}
\end{equation}
where $\mathcal{C}(\qt)$ is the medium collision rate (see e.g.\,\cite{Aurenche:2002pd} in QCD or \cite{Caron-Huot:2006pee,Ghiglieri:2018ltw} in $\mathcal{N}=4$ SYM). At sufficiently large $\kt^2\gg\mu^2$ where $\mu$ is a transverse momentum scale of the order of the Debye screening mass for a hot medium, one can expand $\sigma_{\rm dip}(\rt)$ to leading "twist":
\begin{equation}
\sigma_{\rm dip}(\rt)=\frac{1}{4}\qhat(1/\rt^2,L)\rt^2+\mathcal{O}(\rt^4)\,.\label{eq:qhat-def}
\end{equation}
This relation defines the quenching parameter $\qhat$. At leading order, $\qhat$ depends logarithmically on $\rt$, $\qhat_{\rm LO}=\qhat_0\ln(1/(\rt^2\mu^2))$ because of the Coulomb scattering contribution to $\mathcal{C}(\qt)$. In $\mathcal{N}=4$ SYM, neglecting the hard modes with $\qt\gtrsim T$, one finds $\qhat_0=\lambda Tm_D^2/(4\pi)$ and $m_D^2=2\lambda T^2$ for the Debye screening mass \cite{Caron-Huot:2006pee} (see \cite{Ghiglieri:2018ltw} for the full $\mathcal{O}(\sqrt{\lambda})$ result).  Anticipating our discussion of the resummation of radiative corrections, we have also included a dependence on the system size, although $\qhat$ does not depend on $L$ at leading order for a static medium.

Combining Eqs.\,\eqref{eq:S-matrix} and \eqref{eq:qhat-def}, one observes an emergent transverse momentum scale known as the saturation scale $Q_s$  such that the argument of the exponential is of order one:
\begin{equation}
Q_s^2\equiv\qhat(Q_s^2, L) L\,.\label{eq:Qs-def}
\end{equation}
This saturation scale characterizes the peak of the TMB distribution, as well as the typical $\kt^2$ acquired from the medium (defined using the mediane of the distribution or fractional moments). It allows to distinguish two regimes depending on the value of $k_\perp$ with respect to $Q_s$. For $k_\perp\gg Q_s$, the TMB distribution is dominated by a single hard scattering and the distribution exhibits a typical Rutherford power law decay in $1/\kt^4$ at leading order. For $k_\perp \lesssim Q_s$ one enters into the multiple soft scattering regime and the distribution behaves like a Gaussian. This gaussian approximation justifies the brownian motion in transverse plane paradigm for TMB in the multiple soft scattering regime. Using the leading order estimate $\qhat_{\rm LO}$, one finds that $Q_s^2=\qhat_0 L\ln(\qhat_0L/\mu^2)+\mathcal{O}(\ln\ln L)$. The weak logarithmic dependence aside, the scaling $Q_s^2\sim\qhat_0 L$ corresponds to the standard diffusion scaling. We then define $\gamma(\abar)$ or $\gamma(\lambda)$ as the exponent of $L$ that appears in the large $L$ limit of $Q_s(L)$
\begin{equation}
Q_s(L)\propto L^{\gamma(\lambda)}\,.
\end{equation}
At leading order, the previous discussion gives $\gamma(\lambda)=1/2$.

As mentioned in the introduction, the dominant term in the NLO correction to $\qhat$ in QCD has the double logarithmic form \cite{Liou:2013qya}
\begin{align}
\qhat_{\rm NLO}(\kt^2,L)=\abar\int_{\tau_0}^L\frac{\rmd \tau}{\tau}\int_{Q_s^2(\tau)}^{\kt^2}\frac{\rmd \kt'^2}{\kt'^2}\qhat_{\rm LO}+\mathcal{O}(\alpha_s\ln)\,, \label{eq:qhat-NLO}
\end{align}
where $\tau$ and $\kt'$ are respectively the lifetime and transverse momentum of the gluon fluctuation. For simplicity, we will use $\tau_0=\mu^2/\qhat_0$ as our thermal scale.
In $\mathcal{N}=4$ SYM, one can simply make the replacement $\abar\to \lambda/(4\pi^2)$. The lower boundary for the transverse momentum of the radiated gluon is the saturation momentum so that this emission is typically triggered by a single hard scattering. Using the linear approximation $Q_s(\tau)\approx\qhat_0\tau$, the double integral gives $\qhat_{\rm NLO}(Q_s^2,L)L=\qhat_0/2\times \ln^2(L/\tau_0)$ where the $1/2$ factor is a consequence of the saturation boundary.
The $\mathcal{O}(\alpha_s \ln)$ term refers to single logarithmic corrections which have been computed in \cite{Liou:2013qya,Arnold:2021mow,Arnold:2021pin}.

Beyond NLO, the double logarithmic corrections are resummed through a nonlinear evolution equation that simply iterates Eq.\,\eqref{eq:qhat-NLO} \cite{Iancu:2014kga}. In logarithmic variables $Y=\ln(L/\tau_0)$ and $\rho=\ln(\kt^2/\mu^2)$, this equation reads
\begin{equation}
\frac{\partial \qhat(\rho,Y)}{\partial Y}=\bar\alpha_s\int_{\rho_s(Y)}^\rho\rmd \rho'\,\qhat(\rho',Y)\,,\label{eq:fc-evol}
\end{equation}
where $\rho_s(Y)=\ln(Q_s^2/\mu^2)$.
This equation seems similar to both DGLAP and BFKL equations within the double logarithmic approximation. However, it differs from them in two crucial aspects: (i) the lower boundary of the $\rho'$ integration is bounded from below by the saturation momentum,
(ii) since $\rho_s$ depends on $\qhat$ by definition (see Eq.\,\eqref{eq:Qs-def}), this equation is non-linear. For this reason, it is complicated to solve it analytically without any approximation.

A possible approximation consists in linearizing this equation by using again $Q_s(\tau)\approx\qhat_0\tau \Leftrightarrow\rho_s =Y$ in the lower boundary of the $\rho'$ integral. The evolution equation \eqref{eq:fc-evol} can now be solved analytically for the $\qhat_{\rm LO}$ initial condition \cite{Iancu:2014sha,Mueller:2016xoc}. It is more enlightening to write the result for $Q_s$, defined in the linear case as $Q_s^2(L)=\qhat(\qhat_0 L, L)L$:
\begin{equation}
Q_s^2(L)=\qhat_{\rm LO}L\times \frac{\textrm{I}_1(2\sqrt{\abar Y^2})}{\sqrt{\abar Y^2}}\,.
\end{equation}
Expanding the Bessel function for large $Y$ (large $L$), one easily reads the anomalous exponent: $Q_s^2(L)\propto L^{1+2\sqrt{\abar}}$.
The $\sqrt{\abar}$ term is the one-loop correction to $\gamma(\abar)$. The presence of the square root is a consequence of the double logarithmic nature of the evolution.

If one wants to compute the next term in the development of $\gamma$, which should be of order $\abar$, one needs \textit{a priori} to include in the evolution single logarithmic corrections to the one-loop evolution kernel, the double logarithmic two-loops kernel and also the effects of the non-linear saturation boundary. Before addressing the first two corrections, let us check that indeed, the non-linearity of the evolution equation brings a non-trivial $\mathcal{O}(\abar)$ correction to $\gamma$. We follow the argument presented in \cite{Iancu:2014sha}, and we will check in the next section that our method enables to recover and extend this result to higher orders. If one wants to include the feed-back of the quantum evolution of $Q_s$ in the evolution of $\qhat$ in the large $L$ limit, it is reasonable to use $\rho_s(Y)=(1+2\sqrt{\abar})Y$ in the lower boundary of the $\rho'$ integral in Eq.\,\eqref{eq:fc-evol} instead of $\rho_s=Y$. With this modification, one can still solve analytically the equation, and the evaluation $\qhat$ along the new saturation line gives $Q_s^2(L)\propto L^{1+2\sqrt{\abar(1+2\sqrt{\abar}})}$. Expanding the anomalous exponent in powers of $\abar$, one gets $\gamma(\abar)=1/2+\sqrt{\abar}+\abar+\mathcal{O}(\abar^{3/2})$. As expected, the non-linear behaviour of the evolution equation contributes to the two-loop coefficient of $\gamma$. It is therefore necessary to develop techniques that allow us to address both the non-linear behaviour and the corrections to the double logarithmic kernel of Eq.\,\eqref{eq:fc-evol}. 

\section{Anomalous exponent of the saturation scale at N$^3$LO}
\label{sec:2}
 
A powerful mathematical strategy consists in looking for traveling wave solutions to the evolution equation, as in the case of the asymptotic of the saturation momentum at small Bjorken $x$ \cite{Munier:2003vc,Mueller:2002zm,Beuf:2010aw,Munier:2003sj}. To include the corrections to the kernel beyond the double logarithmic approximation, we write the evolution equation in a BFKL form. Actually, since the evolution is dominated by the double logarithmic regime, as we shall see, one can also use the DGLAP equation \cite{Gribov:1972ri,Altarelli:1977zs,Dokshitzer:1977sg}. In \cite{Caucal:2022mpp}, we show that the two approaches give the same result because of the DGLAP/BFKL duality \cite{Marzani:2007gk}. Let us then write the BFKL equation for the dipole cross-section $\sigma_{\rm dip}(\rho,Y)$ or equivalently for $\qhat$, as
\begin{equation}
\frac{\partial \qhat(\rho,Y)}{\partial Y}=\chi_{\rm BFKL}(\partial_\rho)\left[ \qhat(\rho,Y)\right]\,,\label{eq:BKFL-qhat}
\end{equation}
where the BFKL kernel in Mellin space admits the following expansion $\chi_{BFKL}(\omega)=\abar \chi^{(1)}(\omega)+\abar^2 \chi^{(2)}(\omega)+...$ in planar $\mathcal{N}=4$ SYM ($\abar=\lambda/(4\pi)^2$).

Our traveling wave ansatz to solve this equation in the presence of the saturation boundary takes the form
\begin{equation}
\qhat(\rho,Y)=e^{\rho_s(Y)-Y}e^{\beta x}f(x,Y)\,,
\end{equation}
with $x=\rho-\rho_s(Y)$ and $\beta$ a parameter to be determined. At large $Y$, $f(x,Y)$ converges towards a function $f(x)$ so that this shape corresponds to the propagation along the $\rho$ axis of a front located at $\rho=\rho_s(Y)$. The "time" derivative $\dot\rho_s$ of $\rho_s$ is the front velocity. The $Y$ dependent prefactor $e^{\rho_s(Y)-Y}$ is convenient since the definition \eqref{eq:Qs-def} of the saturation scale simply reads $f(0,Y)=1$ for all $Y$. The consequences of this saturation condition, and especially the additional $e^{-Y}$ factor, are important: contrary to the case of the saturation momentum at small $x$, this extra factor drives the evolution towards the double logarithmic regime of the BFKL equation $\omega\to 0$.

Plugging this ansatz into Eq.\,\eqref{eq:BKFL-qhat}, and expanding the kernel $\chi_{\rm BFKL}(\omega)$ around $\omega=\beta$, we find
\begin{equation}
(\dot\rho_s-1-\dot\rho_s\beta)f-\dot\rho_s \partial_x f+\partial_Y f=\sum\limits_{p=0}^\infty\frac{\chi_{\rm BFKL}^{(p)}(\beta)}{p!}\partial_x^pf\,,\label{eq:fc-LL-exp}
\end{equation}
The existence of a scaling limit $f(x,Y)\to f(x)$ at large time provides two constraints for the large time limit of the front velocity $
c=\lim\limits_{Y\to\infty}\dot\rho_s(Y)$,
related to our anomalous exponent by $c=2\gamma$, and the critical value $\beta_c$ of $\beta$:
\begin{align}
c-1-c\beta_c&=\chi_{\rm BFKL}(\beta_c)\,,\label{eq:1cbeta}\\
-c&=\chi_{\rm BFKL}'(\beta_c)\,.\label{eq:2cbeta}
\end{align}
These relations are obtained after identification of the terms proportional to $f$ and $\partial_x f$ in the $Y\to \infty$ limit of Eq.\,\eqref{eq:fc-LL-exp}.

We first check that these two equations are consistent with the findings of the previous section. The double logarithmic approximation consists in using $\chi_{\rm BFKL}(\omega)\simeq \frac{\abar}{\omega}$ since formally, $\int \rmd\rho =1/\partial_\rho$.
In that case, the system \eqref{eq:1cbeta}-\eqref{eq:2cbeta} can be solved exactly and gives
\begin{align}
c&= 1+2\sqrt{\abar+\abar^2}+2\abar=1+2\sqrt{\abar}+2\abar+\mathcal{O}(\abar^{3/2})\,,\label{eq:c-DLA}
\end{align}
which agrees with the calculation in section \ref{sec:1}. Our method correctly accounts for the non-linear effects in the evolution equation. Another lesson of our approach is that one would obtain exactly the same result by using the \textit{full} leading order BFKL kernel. 
This calculation gives
the same result as Eq.\,\eqref{eq:c-DLA}, the deviation entering at order $\abar^2$ only. This property follows for the dominance of soft and collinear physics in the quantum evolution of the saturation momentum which pushes the "saddle point" $\beta_c\simeq\sqrt{\abar}$ close to $0$ at weak coupling.

One can now take advantage of the recent developments in the computation of the BFKL kernel at three loops in planar $\mathcal{N}=4$ SYM theory \cite{Gromov:2015vua,Caron-Huot:2016tzz,Kotikov:2000pm,Velizhanin:2015xsa}. In fact, the knowledge of the full kernel is not necessary, since we mainly probe the behaviour of the kernel close to $\omega=0$ in the weak coupling limit. To control the $\abar^2$ term, it is sufficient to use
\begin{align}
\chi_{\rm BFKL}(\omega)&=\abar\left(\frac{1}{\omega}+2\zeta(3)\omega^2+\mathcal{O}(\omega^4)\right)+\abar^2\left(\frac{3}{2}\zeta(3)+\mathcal{O}(\omega)\right)\nonumber\\
&+\abar^3\left(-\frac{\zeta(2)}{\omega^3}-\frac{9\zeta(3)}{4\omega^2}+\mathcal{O}\left(\frac{1}{\omega}\right)\right)+\abar^4\left(\frac{k\zeta(3)}{\omega^4}+\mathcal{O}\left(\frac{1}{\omega^3}\right)\right)+\mathcal{O}(\abar^5)\,.
\end{align}
Here, $k$ is a rational number that could be extracted for the BFKL equation at four loops (maybe from \cite{Velizhanin:2021bdh}?)
 Note that one has to systematically subtract the leading collinear poles since those poles are already taken into account by a suitable modification of the leading order kernel \cite{Salam:1998tj,Ciafaloni:1999yw, Altarelli:2005ni,Beuf:2014uia,Ducloue:2019ezk}. 
Solving perturbatively the system \eqref{eq:1cbeta}-\eqref{eq:2cbeta}, our final result for the anomalous exponent reads
\begin{equation}
\gamma(\abar)=\frac{1}{2}+\abar^{1/2}+\abar+\frac{1}{2}\left(1-\zeta(2)\right)\abar^{3/2}+\left(\frac{5+4k}{8}\zeta(3)-2\zeta(2)\right)\abar^2+\mathcal{O}\left(\abar^{5/2}\right)\,,
\end{equation}
with $\zeta$ the Riemann $\zeta$-function ($\zeta(2)=\pi^2/6$ and $\zeta(3)\simeq 1.20206$). The first fours terms of this expression have been computed in \cite{Caucal:2022mpp}.
A few comments are in order. The $\mathcal{O}(\abar)$ is identical to the one derived in the previous section. This is a specificity of the conformal $\mathcal{N}=4$ SYM theory because in QCD, this coefficient would receive a contribution proportional to $\beta_0$ from hard collinear splittings. The $1$ term in the coefficient of $\abar^{3/2}$ is also a consequence of the non-linearity of the evolution, since it already appears at DLA with a saturation boundary. In \cite{Caucal:2022mpp}, we have checked that the first three terms in this expansion can be obtain using the DGLAP evolution equation with a saturation boundary. It would be interesting to cross-check that the following two terms can also be derived from the DGLAP side and DGLAP/BFKL duality \cite{Marzani:2007gk}.

\begin{figure}[h]
\centering
\includegraphics[width=7cm,clip,page=1]{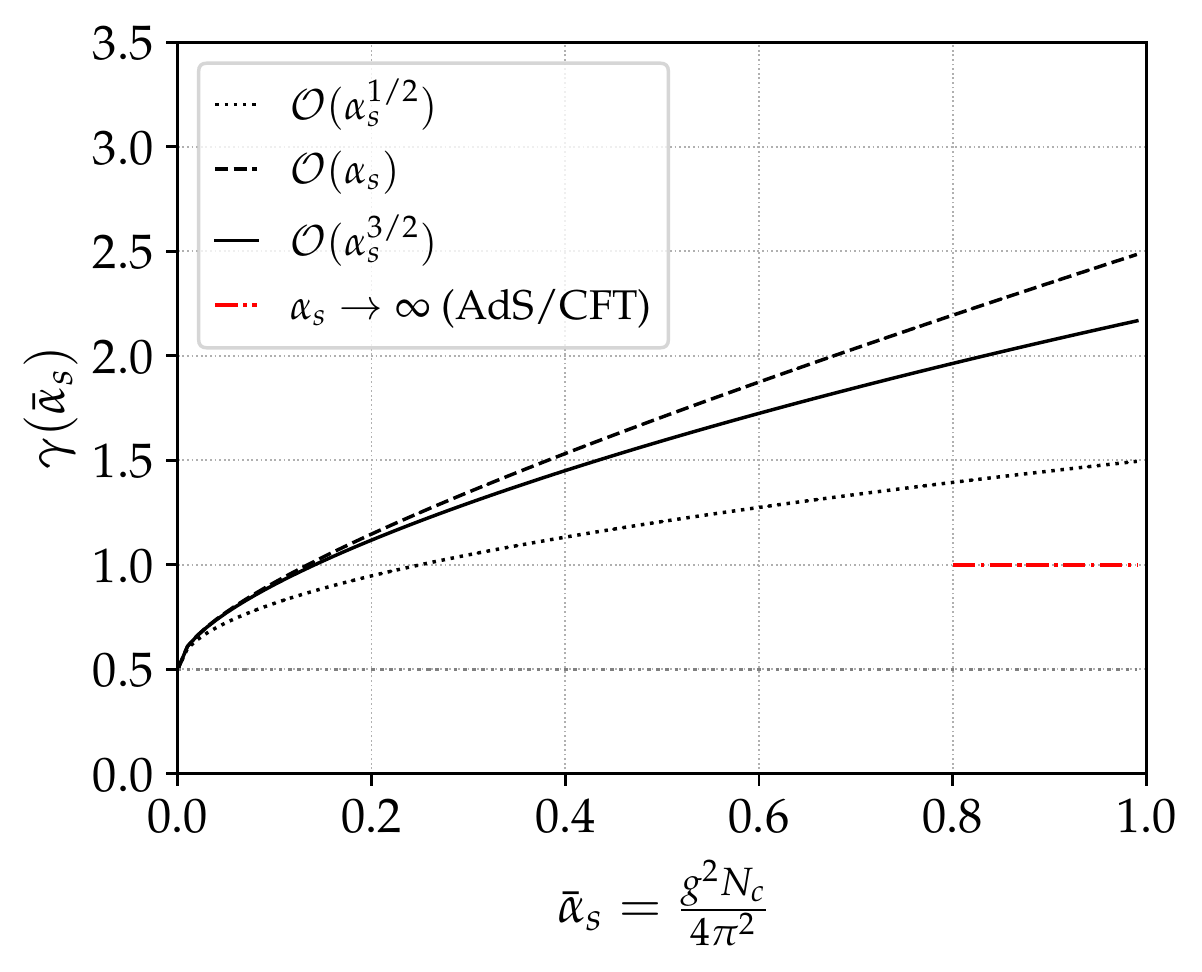}
\caption{Anomalous exponent of transverse momentum diffusion as a function of the 't Hooft coupling for several truncations of the perturbative series. The red curve is the strong coupling limit.}
\label{fig-1}       
\end{figure}

In Figs\,\ref{fig-1}, we display the anomalous exponent $\gamma(\abar)$ as a function of $\abar=\lambda/(4\pi^2)$ at NLO, N$^2$LO and N$^3$LO.  The series 
seems to be convergent, however the $\mathcal{O}(\abar^{3/2})$ curve is above the strong coupling limit $\gamma(\abar\to\infty)=1$ for $\abar\approx 0.2$. This indicates that one should not trust this development beyond this value. Indeed, if the strong coupling limit is correct, the existence of a finite $\abar$ value such that $\gamma(\abar)>1$ implies the existence of an extremum at a critical $\bar{\alpha}_{s}^{c}$. Since there is a priori no reason for the emergence of such a critical value for the dimensionless coupling $\lambda$, we must acknowledge that the series is not reliable beyond $\abar\approx  0.1\div 0.15$. Based on Fig.\,\ref{fig-1}, the range of validity of the weak coupling approach would be rather narrow and $\gamma(\lambda)$ would converge quite rapidly towards the strong coupling limit. This is a striking consequence of our computation. It would be extremely interesting to compute the corrections to the strong coupling limit $\gamma=1$ to confirm this result.


Besides the $\lambda$ development of $\gamma$ (or $c$), the traveling wave method enables one to determine the deviation to $\dot\rho_s=c$ at moderate values of $Y=\ln(L/\tau_0)$ \cite{2003,Munier:2009pc}. The large $Y$ development of the front wave velocity reads \cite{Caucal:2021lgf,Caucal:2022mpp}
\begin{equation}
\dot\rho_s(Y)=2\gamma(\abar)+\frac{3}{2(\beta_c-1)}\frac{1}{Y}+\frac{3}{2(\beta_c-1)^2}\sqrt{\frac{2\pi}{\chi_{\rm BFKL}''(\beta_c)}}\frac{1}{Y^{3/2}}\nonumber\\
+\mathcal{O}\left(Y^{-2}\right)\,,\label{eq:rhos_nonlinear_fixed}
\end{equation}
where each coefficient of the $Y$ power should be expanded in powers of $\abar$ up to the targeted accuracy. This development is \textit{universal}, it does not depend on the initial condition used to solve the full non-linear evolution equation. A consistency check of this universality consists in varying the non-perturbative scale $\tau_0$, which amounts to shift $Y$ by some constant. One easily sees that it does not change the three terms in Eq.\,\eqref{eq:rhos_nonlinear_fixed} but only affects the $1/Y^2$ correction.

\section{Conclusion}

In this paper, we have computed the anomalous exponent $\gamma(\abar)$ up to order $\abar^{2}$ in planar $\mathcal{N}=4$ SYM, based on a pQCD insight. The decisive steps in this calculation are based on two arguments that tremendously simplify the problem: the non-linear evolution of transverse momentum broadening in a dense medium is driven by the double logarithmic approximation and the traveling wave interpretation of the solutions to the evolution equations allows us to use standard techniques developed for the study of reaction-diffusion processes. 

Thanks the the weak coupling development of $\gamma(\abar)$ at three loops, we have discussed the transition towards the strong coupling regime. Our study is similar in spirit to the calculation of the Pomeron intercept from weak to strong coupling done in \cite{Brower:2006ea,Stasto:2007uv,Costa:2012cb,Kotikov:2013xu}. Contrary to the Pomeron intercept case, we conclude that the convergence of the anomalous exponent of the saturation momentum towards the $\lambda\to\infty$ limit $\gamma=1$ should be fast in planar $\mathcal{N}=4$ SYM, typically around $\lambda\approx 4\pi^2/10$. This fact deserves complementary theoretical or numerical studies to be definitively confirmed.

\bibliography{biblio}

\end{document}